\documentclass[10pt,letterpaper]{article}
\usepackage{opex3}
\usepackage{cite}

\begin{document}

\title{Remote transfer of Gaussian quantum discord}
\author{Lingyu Ma and Xiaolong Su$^*$}
\address{State Key Laboratory of Quantum Optics and Quantum Optics Devices,\\ Institute of Opto-Electronics, Shanxi University, Taiyuan, 030006, People's Republic
of China}
\email{$^*$suxl@sxu.edu.cn}

\begin{abstract}
Quantum discord quantifies quantum correlation between quantum systems,
which has potential application in quantum information processing. In this
paper, we propose a scheme realizing the remote transfer of Gaussian quantum
discord, in which another quantum discordant state or an
Einstein-Podolsky-Rosen entangled state serves as ancillary state. The
calculation shows that two independent optical modes that without direct
interaction become quantum correlated after the transfer. The output
Gaussian quantum discord can be higher than the initial Gaussian quantum
discord when optimal gain of the classical channel and the ancillary state
are chosen. The physical reason for this result comes from the fact that the
quantum discord of an asymmetric Gaussian quantum discordant state can be
higher than that of a symmetric one. The presented scheme has potential
application in quantum information network.
\end{abstract}

%% email address is required

%%%%%%%%%%%%%%%%%%% abstract and OCIS codes %%%%%%%%%%%%%%%%
%% [use \begin{abstract*}...\end{abstract*} if exempt from copyright]

% REPLACE WITH CORRECT OCIS CODES FOR YOUR ARTICLE

%%%%%%%%%%%%%%%%%%%%%%% References %%%%%%%%%%%%%%%%%%%%%%%%%

%%%%%%%%%%%%%%%%%%%%%%%%%%  body  %%%%%%%%%%%%%%%%%%%%%%%%%%

\section{Introduction}

Quantum correlation, which can be quantified by quantum discord \cite%
{Ollivier2001,Hen,Modi}, is a fundamental resource for quantum information
processing. It has been shown that quantum correlation (no quantum
entanglement) can be used to complete some quantum computation tasks \cite%
{Knill1998,Ryan2005,Lanyon2008}. Quantum discord can be also utilized as a
resource for entanglement distribution \cite{Stre}, quantum state merging
\cite{Cav,Mad}, remote state preparation \cite{Da}, local broadcasting \cite%
{Pia,Luo} and quantum key distribution \cite{Pir,Su}. Recent research shows
that quantum discord bounds the amount of distributed entanglement \cite%
{Chuan}, cannot be shared \cite{S}, and can be frozen in the non-Markovian
dynamics \cite{OH}. Presently, Quantum discord has been extended to the
region of continuous variables for Gaussian states \cite%
{Giorda2010,Adesso2010} and certain non-Gaussian states \cite{Tat}. Gaussian
quantum discord has been experimentally demonstrated \cite%
{Gu2012,Blandino2012,Madsen2012,Vogl2013}.

Entanglement swapping \cite{Zuk,Tan,Loock1,Zhang1}, which makes two
independent quantum states without direct interaction become entangled, is
important to build quantum information network \cite{Duan}. In fact, it
represents the quantum teleportation of entangled state \cite{Loock1}.
Entanglement swapping has been experimentally demonstrated in both discrete
and continuous variables region \cite{Pan,Jia}.

In this paper, we propose a scheme to realize the remote and unconditional
transfer of Gaussian quantum discord based on applying the technique of
entanglement swapping. The Gaussian quantum discord is transferred remotely
and unconditionally by using an other quantum discordant state or an
Einstein-Podolsky-Rosen (EPR) entangled state as ancillary state. Two
quantum states that never have direct interaction become quantum correlated
after transfer of Gaussian quantum discord. A more interesting result is
that the output Gaussian quantum discord can be higher than the initial
quantum discord at some given conditions, which will never happen in
entanglement swapping. The maximum output quantum discord is obtained when
optimal gain in classical channel and squeezing parameter of the EPR
entangled state are chosen. The simplest way to transfer the Gaussian
quantum discord is to use a coherent state as ancillary state, where the
output quantum discord can also be higher than the initial quantum discord
when optimal gain is chosen. This feature is useful for constructing a
quantum information network. The physical reason for this result is that the
quantum discord of an asymmetric Gaussian quantum discordant state can be
higher than that of a symmetric one, which is also proved in the paper.
Here, the asymmetric (symmetric) Gaussian quantum discordant state means
that the correlated noise variances on two quantum correlated beams are
different (same).

\section{The remote transfer scheme}

\subsection{Using a Gaussian discordant state as ancillary state}

Fig. 1 shows the schematic for the remote transfer of Gaussian quantum
discord. Alice and Bob, who represent two nodes in a quantum information
network, own two independent Gaussian discordant states ($\hat{a},\hat{b}$)
and ($\hat{c},\hat{d}$), respectively. In order to transfer the quantum
correlation to remote station, Alice and Bob send one of each states ($\hat{b%
}$ and $\hat{c}$) to the middle station owned by Claire. Claire interferes
quantum modes $\hat{b}$ and $\hat{c}$ on a 50\% beam-splitter (BS) and
measures the amplitude and phase quadratures of the output modes $\hat{f}$\
and $\hat{e}$ by two homodyne detection (HD) systems, respectively. The
measurement results are fedforward to Bob through the classical channel. Bob
performs phase space displacement on quantum mode $\hat{d}$ with amplitude
and phase modulators (EOMx and EOMp), respectively. Finally, the quantum
discord between quantum modes $\hat{a}$ and $\hat{d}%
%TCIMACRO{\U{b4}}%
%BeginExpansion
{\acute{}}%
%EndExpansion
$ are measured by an verifier.

\begin{figure}[tbp]
\setlength{\belowcaptionskip}{-3pt}
\centerline{
\includegraphics[width=100mm]{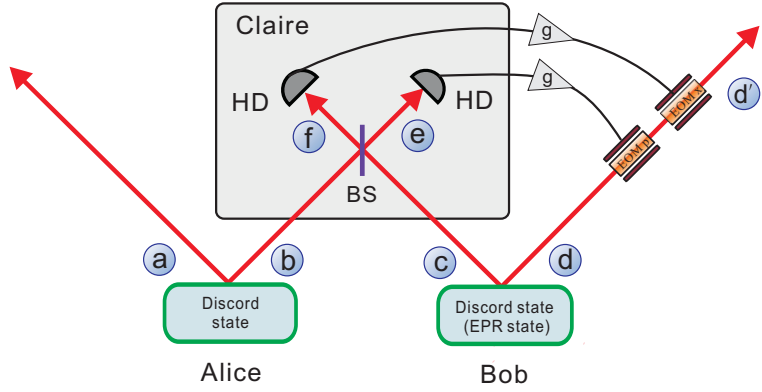}
}
\caption{Schematic of the Gaussian quantum discord remote transfer. BS:
50:50 beam-splitter, HD: homodyne detection system, g: gain factor of the
classical channel, EOMx and EOMp: amplitude and phase electro-optical
modulator.}
\label{Fig1}
\end{figure}

The amplitude and phase quadratures of an optical mode $\hat{o}$ are defined
as $\hat{X}_{o}=\hat{o}+\hat{o}^{\dagger }$ and $\hat{Y}_{o}=(\hat{o}-\hat{o}%
^{\dagger })/i$, respectively. The variances of amplitude and phase
quadratures for a vacuum (coherent) state are $V(\hat{X}_{v})=V(\hat{Y}%
_{v})=1$. The Gaussian quantum discordant state can be obtained by
correlated (anti-correlated) displacement of two coherent states in the
amplitude (phase) quadrature with a discording noise $V$ \cite{Gu2012},
which is a separable state. The two modulated coherent states are $\hat{a}=%
\hat{c}_{1}+\hat{s}_{1}$ and $\hat{b}=\hat{c}_{2}+\hat{s}_{2}$,
respectively, where $\hat{c}_{1}$ and $\hat{c}_{2}$ represent two
independent coherent state at same frequency, $\hat{s}_{1}=\hat{X}_{s}+i\hat{%
Y}_{s}$ and $\hat{s}_{2}=\hat{X}_{s}-i\hat{Y}_{s}$ stand for the discording
signal with a variance $V(\hat{X}_{s})=V(\hat{Y}_{s})=V$. The covariance
matrix of the prepared Gaussian discordant state is given by
\begin{equation}
\ \mathbf{\sigma }=\left(
\begin{array}{cc}
\mathbf{A} & \mathbf{C} \\
\mathbf{C} & \mathbf{B}%
\end{array}%
\right) ,
\end{equation}%
where
\begin{equation}
\mathbf{A}=\mathbf{B}=\left(
\begin{array}{cc}
1+V & 0 \\
0 & 1+V%
\end{array}%
\right) ,\qquad \mathbf{C}=\left(
\begin{array}{cc}
V & 0 \\
0 & -V%
\end{array}%
\right) .
\end{equation}%
For Alice and Bob's Gaussian discordant state, we have $V=V_{A}$ and $%
V=V_{B} $ in the covariance matrix, respectively.

Quantum discord is defined as the difference between two quantum analogues
of classically equivalent expression of the mutual information. For a
bipartite quantum system $\rho _{AB}$, the total classical and quantum
correlations is given by mutual information $I(\rho _{AB})=S(\rho
_{A})+S(\rho _{B})-S(\rho _{AB})$, where $S(\rho )$ is the von Neumann
entropy. Another expression of mutual information based on conditional
entropy is $J^{\leftarrow }(\rho _{AB})=S(\rho _{A})-\inf_{\{\Pi
_{j}\}}S_{\{\Pi _{j}\}}(A|B)$, which is known as one-way classical
correlation. The infimum describes all possible measurements on subsystem $B$%
. The difference between these two expressions of mutual information is
defined as quantum discord $D_{AB}=I(\rho _{AB})-J^{\leftarrow }(\rho _{AB})$%
. An explicit expression of Gaussian quantum discord for a two-mode Gaussian
state is given by \cite{Adesso2010}
\begin{equation}
D_{AB}=f(\sqrt{I_{2}})-f(\nu _{-})-f(\nu _{+})+f(\sqrt{E^{\min }}),
\label{discord}
\end{equation}%
where $f(x)=(\frac{x+1}{2})\log _{2}({\frac{x+1}{2})}-(\frac{x-1}{2})\log
_{2}({\frac{x-1}{2})}${, }%
\begin{equation}
\nu _{\pm }=\sqrt{\frac{\Delta \pm \sqrt{\Delta ^{2}-4\det \mathbf{\sigma }}%
}{2}}
\end{equation}%
are the symplectic eigenvalues of a two-mode covariance matrix $\mathbf{%
\sigma }${\ with }$\det \mathbf{\sigma }$ as the determinant of covariance
matrix and $\Delta =\det \mathbf{A}+\det \mathbf{B}+2\det \mathbf{C}$, and
\begin{equation}
E^{\min }=\left\{
\begin{array}{ll}
\frac{2I_{3}^{2}+(I_{2}-1)(I_{4}-I_{1})+2|I_{3}|\sqrt{%
I_{3}^{2}+(I_{2}-1)(I_{4}-I_{1})}}{(I_{2}-1)^{2}}\quad & a) \\
\frac{I_{1}I_{2}-I_{3}^{2}+I_{4}-\sqrt{%
I_{3}^{4}+(I_{4}-I_{1}I_{2})^{2}-2I_{3}^{2}(I_{4}+I_{1}I_{2})}}{2I_{2}}\quad
& b)%
\end{array}%
\right.
\end{equation}%
where a) applies if $\quad (I_{4}-I_{1}I_{2})^{2}\leq
I_{3}^{2}(I_{2}+1)(I_{1}+I_{4})$ and b) applies otherwise. $I_{1}=\det
\mathbf{A}$, $I_{2}=\det \mathbf{B}$, $I_{3}=\det \mathbf{C}$, $I_{4}=\det {%
\sigma }$ are the symplectic invariants. A two-mode Gaussian state is
quantum correlated when Gaussian quantum discord $D_{AB}>0$.

In Fig. 1, the output modes from the 50\% beam-splitter are $\hat{e}=(\hat{b}%
+\hat{c})/\sqrt{2}$\ and $\hat{f}=(\hat{b}-\hat{c})/\sqrt{2}$. The measured
photocurrents of two homodyne detection systems are $\hat{\imath}_{1}=(\hat{X%
}_{b}-\hat{X}_{c})/\sqrt{2}$ and $\hat{\imath}_{2}=(\hat{Y}_{b}+\hat{Y}_{c})/%
\sqrt{2}$, respectively. They are sent to Bob through the classical channel,
respectively. The output beam is%
\begin{equation}
\hat{d}^{\prime }=\hat{d}+\sqrt{2}g\hat{\imath}_{1}+i\sqrt{2}g\hat{\imath}%
_{2},
\end{equation}%
where $g$ describes Bob's (suitably normalized) amplitude and phase gain
factor in the classical channels, where we have assumed that the gain in two
channels is identical. The covariance matrix of the output states $\hat{a}$
and $\hat{d}%
%TCIMACRO{\U{b4}}%
%BeginExpansion
{\acute{}}%
%EndExpansion
$ is given by
\begin{equation}
\sigma _{out}=\left(
\begin{array}{cc}
\mathbf{A%
%TCIMACRO{\U{b4}}%
%BeginExpansion
{\acute{}}%
%EndExpansion
} & \mathbf{C%
%TCIMACRO{\U{b4}}%
%BeginExpansion
{\acute{}}%
%EndExpansion
} \\
\mathbf{C%
%TCIMACRO{\U{b4}}%
%BeginExpansion
{\acute{}}%
%EndExpansion
} & \mathbf{B%
%TCIMACRO{\U{b4}}%
%BeginExpansion
{\acute{}}%
%EndExpansion
}%
\end{array}%
\right)  \label{out}
\end{equation}%
where%
\[
\mathbf{A%
%TCIMACRO{\U{b4}}%
%BeginExpansion
{\acute{}}%
%EndExpansion
}\mathbf{=}\left(
\begin{array}{cc}
1+V_{A} & 0 \\
0 & 1+V_{A}%
\end{array}%
\right) ,\mathbf{\quad \mathbf{B%
%TCIMACRO{\U{b4}}%
%BeginExpansion
{\acute{}}%
%EndExpansion
=}}\left(
\begin{array}{cc}
V_{d%
%TCIMACRO{\U{b4}}%
%BeginExpansion
{\acute{}}%
%EndExpansion
} & 0 \\
0 & V_{d%
%TCIMACRO{\U{b4}}%
%BeginExpansion
{\acute{}}%
%EndExpansion
}%
\end{array}%
\right) ,\mathbf{\quad }\mathbf{C%
%TCIMACRO{\U{b4}}%
%BeginExpansion
{\acute{}}%
%EndExpansion
}=\left(
\begin{array}{cc}
gV_{A} & 0 \\
0 & -gV_{A}%
\end{array}%
\right) ,
\]%
with $V_{d%
%TCIMACRO{\U{b4}}%
%BeginExpansion
{\acute{}}%
%EndExpansion
}=1+2g^{2}+g^{2}V_{A}+(1-g)^{2}V_{B}$. Finally, the quantum discord of the
output states can be verified by Eq. (\ref{discord}).

\subsection{Using an EPR entangled state as ancillary state}

For the case of transferring Gaussian quantum discord with an EPR
entangled state as ancillary state, the quantum discordant state at
Bob's station is replaced by an EPR entangled state, whose
covariance matrix is given by
\begin{equation}
\ \mathbf{\sigma }_{E}=\left(
\begin{array}{cc}
\mathbf{A}_{E} & \mathbf{C}_{E} \\
\mathbf{C}_{E} & \mathbf{B}_{E}%
\end{array}%
\right) ,
\end{equation}%
where
\begin{equation}
\mathbf{A}_{E}=\mathbf{B}_{E}=\left(
\begin{array}{cc}
V_{E} & 0 \\
0 & V_{E}%
\end{array}%
\right) ,\quad \mathbf{C}_{E}=\left(
\begin{array}{cc}
\sqrt{V_{E}^{2}-1} & 0 \\
0 & -\sqrt{V_{E}^{2}-1}%
\end{array}%
\right) ,
\end{equation}%
in which $V_{E}=\cosh 2r$ is the variance of the EPR entangled state, where $%
r$ is the squeezing parameter.

After the transfer of the quantum discordant state, the covariance matrix of
the output state becomes
\begin{equation}
\sigma _{Eout}=\left(
\begin{array}{cc}
\mathbf{A%
%TCIMACRO{\U{b4}}%
%BeginExpansion
{\acute{}}%
%EndExpansion
} & \mathbf{C%
%TCIMACRO{\U{b4}}%
%BeginExpansion
{\acute{}}%
%EndExpansion
} \\
\mathbf{C%
%TCIMACRO{\U{b4}}%
%BeginExpansion
{\acute{}}%
%EndExpansion
} & \mathbf{B%
%TCIMACRO{\U{b4}}%
%BeginExpansion
{\acute{}}%
%EndExpansion
}_{E}%
\end{array}%
\right) ,
\end{equation}%
where $\mathbf{A%
%TCIMACRO{\U{b4}}%
%BeginExpansion
{\acute{}}%
%EndExpansion
}$ and $\mathbf{C%
%TCIMACRO{\U{b4}}%
%BeginExpansion
{\acute{}}%
%EndExpansion
}$ are same with that in Eq. (\ref{out}), and
\begin{equation}
\mathbf{B%
%TCIMACRO{\U{b4}}%
%BeginExpansion
{\acute{}}%
%EndExpansion
}_{E}\mathbf{=}\left(
\begin{array}{cc}
V_{Ed%
%TCIMACRO{\U{b4}}%
%BeginExpansion
{\acute{}}%
%EndExpansion
} & 0 \\
0 & V_{Ed%
%TCIMACRO{\U{b4}}%
%BeginExpansion
{\acute{}}%
%EndExpansion
}%
\end{array}%
\right)
\end{equation}%
with $V_{Ed%
%TCIMACRO{\U{b4}}%
%BeginExpansion
{\acute{}}%
%EndExpansion
}=(g^{2}+1)V_{E}-2g\sqrt{V_{E}^{2}-1}+g^{2}(1+V_{A})$.

\subsection{The effect of imperfect detection efficiency}

We also consider the effect of the imperfect detection efficiency on the
proposed transfer schemes. The detection efficiency can be looked as the
loss of the detected optical mode, which is usually modeled by a beam
splitter with transmission efficiency of $\eta $. In this way, the detected
optical mode $\hat{o}$ turns into $\sqrt{\eta }\hat{o}+\sqrt{1-\eta }\hat{\nu%
}$, where $\hat{\nu}$ represent the vacuum noise induced by
imperfect detection efficiency. For analyzing the effect of
imperfect detection efficiency on the proposed scheme, we consider
the detection efficiency on the optical modes $\hat{a}$, $\hat{e}$,
$\hat{f}$, and $\hat{d}^{\prime }$, respectively.

\section{Results and analysis}

\begin{figure}[tbp]
\setlength{\belowcaptionskip}{-3pt}
\centerline{
\includegraphics[width=130mm]{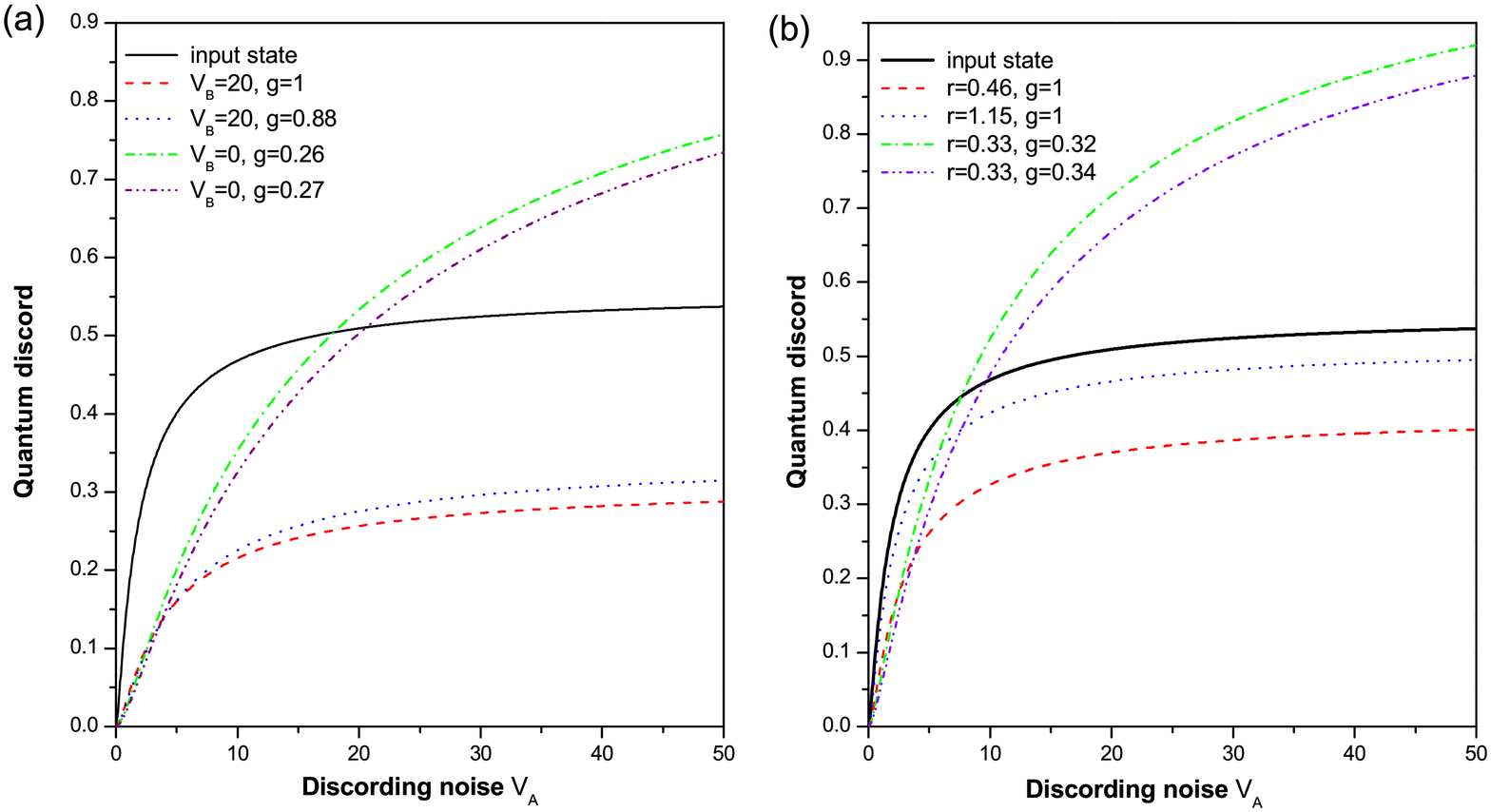}
}
\caption{Dependence of quantum discord of output state on the discording
noise with a quantum discordant state (a) and an EPR entangled state (b) as
ancillary state.}
\label{Fig2}
\end{figure}

Figure 2(a) shows the dependence of output quantum discord on the Alice's
discording noise $V_{A}$ with a Gaussian quantum discordant state with
discording noise $V_{B}$ as ancillary state. The output quantum discord with
$V_{B}=20$, $g=1$ (red dashed line) is lower than the Alice's initial
quantum discord (black solid line). From the expression of $V_{d%
%TCIMACRO{\U{b4}}%
%BeginExpansion
{\acute{}}%
%EndExpansion
}$, we can see that the output quantum discord is independent on Bob's
discording noise when a unit gain factor is chosen. When the optimal gain
factor is chosen, the output quantum discord can be improved (blue dotted
line with $V_{B}=20$, $g=0.88$). The most important result is that the
output quantum discord can be higher than the initial quantum discord when
optimal gain factor and Bob's discording noise are chosen. For example, the
green dash-dotted line in Fig. 2(a) corresponds to $V_{B}=0$, $g=0.26$,
which means that two independent coherent states are used as ancillary
state. It is clear that the output quantum discord is higher than the
initial quantum discord when the discording noise of initial state $V_{A}>18$%
. Therefore we can complete the transfer of Gaussian quantum discord using
coherent state as ancillary state, which is the simplest way for the
practical application.

Figure 2(b) shows the dependence of output quantum discord on the discording
noise with an EPR entangled state as ancillary state. Comparing the red
dashed ($r=0.46$, $g=1$) and blue dotted lines ($r=1.15$, $g=1$), we find
that the higher entanglement the higher output quantum discord can be
obtained when unit gain factor is chosen. When the optimal squeezing and
gain factor are chosen, the output quantum discord (green dash-dotted line
with $r=0.33$, $g=0.32$) can also be higher than the initial quantum discord
(black solid line) when the initial discording noise $V_{A}>7.6$. Comparing
the green dash-dotted lines in Fig. 2(a) and 2(b), we find that the maximum
output Gaussian quantum discord is obtained when an EPR entangled state is
used as ancillary state with optimal parameters. The purple
dash-dotted-dotted lines in Fig. 2 correspond to the case of imperfect
detection efficiency with $\eta =0.9$\ and optimal gain factors. Comparing
the green dash-dotted and purple dash-dotted-dotted lines, it is obvious
that the quantum discord is decreased due to the imperfect detection efficiency.

\begin{figure}[tbp]
\setlength{\belowcaptionskip}{-3pt}
\centerline{
\includegraphics[width=130mm]{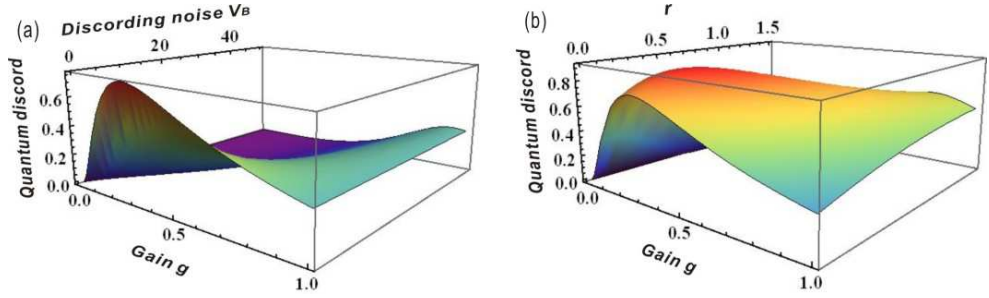}
}
\caption{(a) Dependence of output quantum discord on Bob's discording noise
and the gain factor with a quantum discordant state as ancillary state. (b)
Dependence of output quantum discord on squeezing parameter and the gain
factor with an EPR entangled state as ancillary state. }
\end{figure}

Figure 3(a) shows the dependence of output quantum discord on Bob's
discording noise and gain factor for the case with a quantum discordant
state as ancillary state, where Alice's discording noise is chosen to be 50.
It is obvious that the maximum output quantum discord is obtained at $%
V_{B}=0 $, i.e. the ancillary states are two independent coherent states.
This confirms that using coherent state as ancillary state is the simplest
way to transfer the Gaussian quantum discord. Figure 3(b) shows the
dependence of output quantum discord on squeezing parameter and gain factor
when an EPR entangled state is used as an ancillary state, in which Alice's
discording noise is fixed to 50. The maximum output quantum discord is
obtained with optimal squeezing parameter and gain factor. Comparing the
maximum output quantum discord in Fig. 3(a) and 3(b), we find that
entanglement is helpful to obtain maximum output quantum discord.

\begin{figure}[tbp]
\setlength{\belowcaptionskip}{-3pt}
\centerline{
\includegraphics[width=130mm]{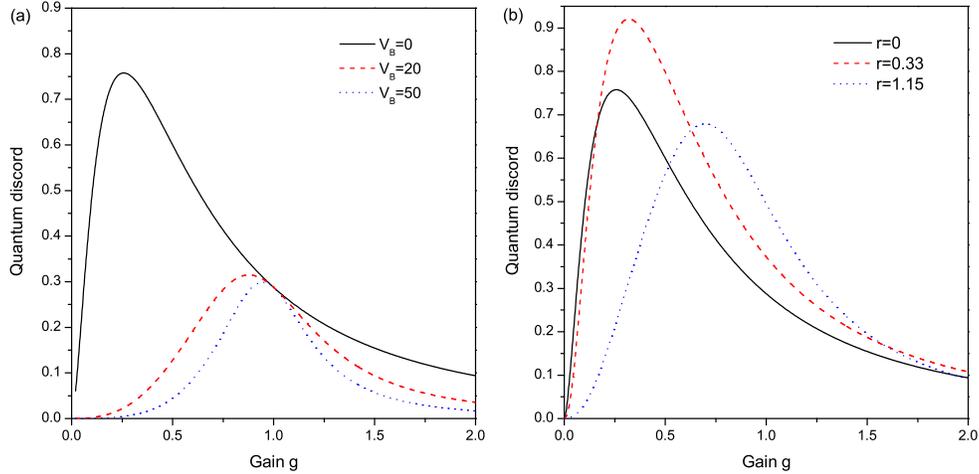}
}
\caption{Dependence of quantum discord of output state on the gain factor
with a quantum discordant state (a) and an EPR entangled state (b) as
ancillary state.}
\end{figure}

The relation between quantum discord of the output states and the gain
factor when Alice's discording noise $V_{A}=50$ is shown in Fig. 4. In Fig.
4(a), the black solid, red dashed and blue dotted lines correspond to Bob's
discording noise of 0, 20 and 50, respectively. The optimal gain factor
increases with the increasing of Bob's discording noise. The maximum output
quantum discord is obtained when coherent state is used as ancillary state.
In Fig. 4(b), the black solid, red dashed and blue dotted lines correspond
to squeezing parameter $r=0,$ $0.33$ and $1.15$, respectively. The optimal
gain factor is close to unit with the increasing of squeezing parameter.
There is an optimal squeezing that maximize the output quantum discord,
which is $r=0.33$. In this case, the maximum output Gaussian quantum discord
of $0.92$ is obtained.

\section{Quantum discord of an asymmetric Gaussian discordant state}

We now analyze the physical reason for why output quantum discord can be
higher than initial quantum discord. Comparing the covariance matrixes of
the initial and output Gaussian discordant states, we find that the initial
state is a symmetric quantum discordant state, which has same noise
variances on its two quantum correlated beams, while the output state is an
asymmetric quantum discordant state, which has different noise variances on
its two quantum correlated beams. This implies that the quantum discord of
an asymmetric Gaussian quantum discordant state may be higher than that of a
symmetric one. For proving this conclusion, we assume the discording noise
modulated on two coherent states are different in the preparation of the
quantum discordant state. For example, assuming the two modulated coherent
states are $\hat{a}=\hat{c}_{1}+\hat{s}_{1}$ and $\hat{b}=\hat{c}_{2}+T\hat{s%
}_{2}$, respectively, where $T$ is the attenuation of the discording signal.
In this case, the elements in the covariance matrix of the prepared Gaussian
quantum discordant state [Eq. (1)] becomes
\begin{equation}
\mathbf{B}=\left(
\begin{array}{cc}
1+T^{2}V & 0 \\
0 & 1+T^{2}V%
\end{array}%
\right) \qquad \mathbf{C}=\left(
\begin{array}{cc}
TV & 0 \\
0 & -TV%
\end{array}%
\right) ,
\end{equation}%
When $T\neq 1$, the prepared state becomes an asymmetric Gaussian quantum
discordant state.

\begin{figure}[tbp]
\setlength{\belowcaptionskip}{-3pt}
\centerline{
\includegraphics[width=130mm]{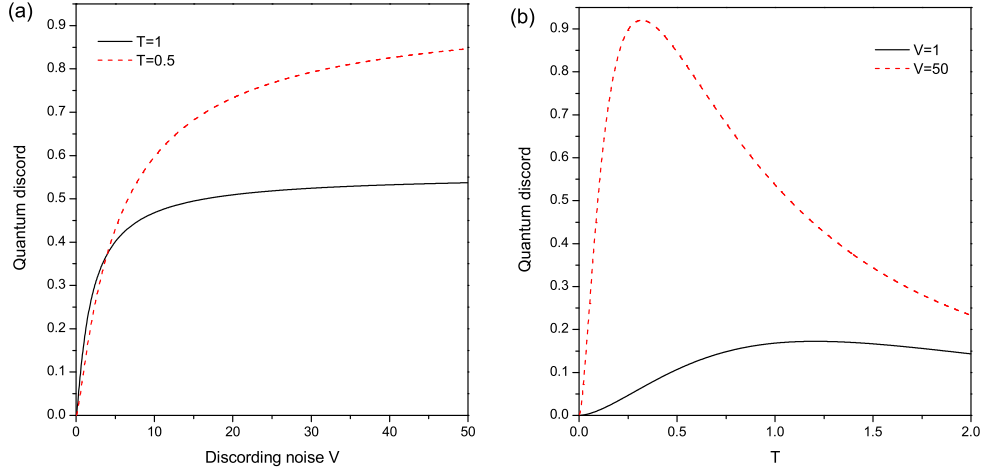}
}
\caption{(a) Quantum discord of the asymmetric ($T=0.5$, red dashed line)
and symmetric ($T=1$, black solid line) Gaussian quantum discordant state.
(b) Dependence of the Gaussian quantum discord on attenuation at discording
noises $V=50$ (red dashed line) and $V=1$ (black solid line).}
\label{Fig5}
\end{figure}

Figure 5(a) shows the Gaussian quantum discord for the prepared state. It is
obvious that the quantum discord with attenuation (red dashed line, $T=0.5$)
is higher than that without attenuation (black solid line, $T=1$) when
discording noise $V>4$. Fig. 5(b) shows the dependence of quantum discord on
the attenuation $T$ at different discording noises. When $V=50$ (red dashed
line), it is clear that there is an optimal attenuation $T$ that maximize
the quantum discord. When $0.1<T<1$, the quantum discord is higher than the
symmetric Gaussian quantum discordant state. However, when $V=1$ (black
solid line), the quantum discord of the asymmetric Gaussian quantum
discordant state is not higher than that of the symmetric one.

A symmetric Gaussian discordant state can also be turned into an
asymmetric state by attenuate one or both optical modes. Here, we
attenuate the optical modes $\hat{a}$ and $\hat{b}$ by two beam
splitters with transmission efficiencies $T_{1}$ and $T_{2}$,
respectively. The attenuated optical modes
become $\hat{a}%
%TCIMACRO{\U{b4}}%
%BeginExpansion
{\acute{}}%
%EndExpansion
=\sqrt{T_{1}}\hat{a}+\sqrt{1-T_{1}}\hat{v}_{1}=\sqrt{T_{1}}(\hat{c}_{1}+\hat{%
s}_{1})+$ $\sqrt{1-T_{1}}\hat{v}_{1}$ and $\hat{b}%
%TCIMACRO{\U{b4}}%
%BeginExpansion
{\acute{}}%
%EndExpansion
=\sqrt{T_{2}}\hat{b}+\sqrt{1-T_{2}}\hat{v}_{2}=\sqrt{T_{2}}(\hat{c}_{2}+\hat{%
s}_{2})+$ $\sqrt{1-T_{2}}\hat{v}_{2}$, respectively.

\begin{figure}[tbp]
\setlength{\belowcaptionskip}{-3pt}
\centerline{
\includegraphics[width=80mm]{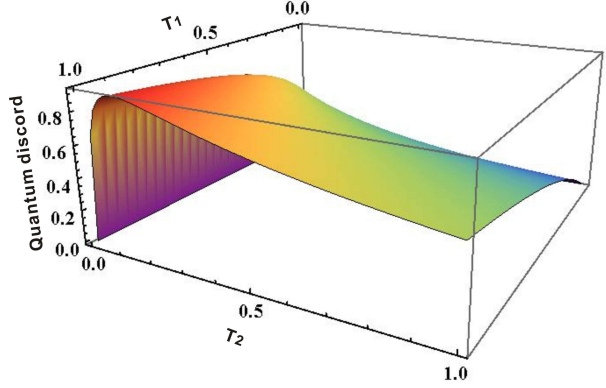}
}
\caption{Dependence of quantum discord on the attenuations $T_{1}$ and $%
T_{2} $ with discording noise $V=50$.}
\end{figure}

Figure 6 shows the dependence of quantum discord on the attenuations $T_{1}$
and $T_{2}$ with discording noise $V=50$. When $T_{1}=1$ and $0<T_{2}<1$,
the Gaussian discordant state is an asymmetric state. It is obvious that the
quantum discord with the optimal attenuation $T_{2}$ is higher than the
discord of a symmetric Gaussian discordant state, which corresponds to the
quantum discord at $T_{1}=T_{2}=1$. This is similar to what we obtained in
Fig. 5(b), which confirms that the quantum discord of an asymmetric Gaussian
discordant state can be higher than a symmetric one with optimal
attenuation. When the attenuation on the mode $\hat{a}$ is also considered,
the quantum discord is decreased along with the decrease of $T_{1}$. This is
because the asymmetry between two optical modes is reduced by attenuating
both modes.

So we draw a conclusion that the quantum discord of an asymmetric Gaussian
quantum discordant state can be higher than that of a symmetric one prepared
in this way when optimal attenuation and discording noise are chosen. This
comes from the fact that the one-way classical correlation $J^{\leftarrow
}(\rho _{AB})$ is measurement dependent. The attenuation of one mode
(asymmetry) lead to the decrease of total and classical correlations, while
the difference between them (quantum discord) can be increased. This result
is compatible with what obtained by Madsen et al in \cite{Madsen2012}, where
the attenuation of one mode of a Gaussian quantum discordant state lead to
increasing of quantum discord. In \cite{Madsen2012}, transmission of one
mode of a Gaussian discordant state in a lossy channel is mimicked by a
beam-splitter with variable transmission efficiency. They observed the
increase of quantum discord along with the increasing of the attenuation,
which lead to an asymmetric Gaussian quantum discordant state. This property
of Gaussian quantum discord will have potential application in quantum
information tasks. For example, in the presented remote transfer scheme, the
output quantum discord can be higher than the input discord. Since the
Gaussian quantum discordant state is resilient to loss, it may be used to
extend the transmission distance of quantum communication.

\section{Conclusion}

In conclusion, a remote transfer scheme of Gaussian quantum discord is
proposed. The quantum correlation emerges between two independent quantum
modes that without direct interaction after the remote transfer. In the
remote transfer scheme with a Gaussian discordant state as ancillary state,
the output quantum discord is independent on discording noise of the
ancillary state when unit gain is chosen. The maximum output quantum discord
is obtained with two independent coherent states as ancillary state in the
case of optimal gain. In the remote transfer scheme with an EPR entangled
state as ancillary state, the maximum output quantum discord is obtained
when optimal squeezing and gain are chosen. Comparing schemes with a
Gaussian quantum discordant state and an EPR entangled state as ancillary
state, we find that entanglement in the ancillary state is helpful to
improve the output quantum discord.

Notably, the output Gaussian quantum discord can be higher than the initial
quantum discord. The physical reason for this phenomenon is that the quantum
discord of an asymmetric Gaussian quantum discordant state can be higher
than that of a symmetric one. The remote transfer of Gaussian quantum
discord has potential application in the quantum information network.
Especially, the coherent states can serve as the ancillary states, which
might save the quantum resources in the practical application.

\section*{ACKNOWLEDGMENTS}

We thank for helpful discussion with Prof. Changde Xie. This research was
supported by the National Basic Research Program of China (Grant No.
2010CB923103), NSFC (Grant Nos. 11174188, 61121064), Shanxi Scholarship
Council of China (Grant No. 2012-010) and OIT.

\end{document}